\documentclass[10pt,letterpaper,twoside,twocolumn,american,nofootinbib,prd,aps]{revtex4}
\usepackage{ae}
\usepackage{aecompl}
\usepackage[T1]{fontenc}
\usepackage[latin1]{inputenc}
\usepackage{amsmath}
\usepackage{graphicx}
\usepackage{amssymb}

\makeatletter
\usepackage{psfrag}

\usepackage{babel}
\makeatother
\begin{document}

\title{Time-reparametrization invariance in eternal inflation }

\author{Sergei Winitzki}

\affiliation{Department of Physics, Ludwig-Maximilians University, Theresienstr.~37, 80333
Munich, Germany}

\begin{abstract}
I address some recently raised issues regarding the time-parametrization dependence
in stochastic descriptions of eternal inflation. To clarify the role of the
choice of the time gauge, I show examples of gauge-dependent as well as gauge-independent
statements about physical observables in eternally inflating spacetimes. In
particular, the relative abundance of thermalized and inflating regions is highly
gauge-dependent. The unbounded growth of the 3-volume of the inflating regions
is found in certain time gauges, such as the proper time or the scale factor
gauge. Yet in the same spacetimes there exist time foliations with a finite
and monotonically decreasing 3-volume, which I demonstrate by an explicit construction.
I also show that there exists no {}``correct\char`\"{} choice of the time gauge
that would yield an unbiased stationary probability distribution for observables
in thermalized regions.
\end{abstract}
\maketitle

\section{Introduction}

In most inflationary scenarios, the exit from inflation and the subsequent reheating
do not occur everywhere at once. Some regions of the universe finish inflation
and thermalize, while other regions continue inflating. Generic models of inflation
predict that such inflating regions will be found at arbitrarily late times.
This phenomenon was called \textsl{eternal inflation}%
\footnote{Since it was not implied that inflation is eternal also to the past, a more
precise designation would be {}``future-eternal inflation.\char`\"{} %
}~\cite{Vil83,Lin86,Sta86}. The inflating regions expand much faster than the
thermalized regions, so heuristically one expects that most of the volume of
the universe is dominated by regions that have been inflating for a very long
time and thus retain no memory of the initial state. The planet Earth is equally
likely to be in any of the (infinitely many) thermalized regions, hence our
present-day observations are essentially independent of the conditions before
the onset of inflation. In this way the picture of eternal inflation was found
to alleviate the problem of initial conditions for inflation. Any initial state
is permissible (agrees with our present observations) as long as there is a
nonzero probability for eternal inflation to set in. 

The inflating (i.e.~not yet thermalized) domain in an eternally inflating spacetime
can be described as a random collection of horizon-sized regions with locally
de Sitter geometry. All regions expand at different local Hubble rates $H\left(t,x\right)$,
where $x$ is a comoving coordinate and $t$ is the proper time measured along
comoving geodesics $x=\textrm{const}$. A region with expansion rate $H$ evolves
during one Hubble time $\Delta t\sim H^{-1}$ into $e^{3}$ new horizon-sized
regions having slightly changed expansion rates; some of these new regions may
accidentally thermalize. Thus the evolution of the spacetime can be viewed as
a realization of a branching diffusion process. One can compute probability
distributions for various physical observables, such as the scalar field $\phi$
or the scale factor $a$, at a given time $t$. Calculations are performed by
modeling the branching diffusion process using Langevin and Fokker-Planck (FP)
equations (see \cite{LinLinMez94} for a review and further references). Below
I shall summarily refer to this description as the \textsl{FP formalism}. 

A description through an equal-time probability distribution necessarily suffers
from a dependence on the choice of the time variable $t$ and in some cases
this gauge dependence is quite dramatic~\cite{LinLinMez94,GarLin95,Vil95}.
The source of the gauge dependence is a bias introduced into the selection of
observers when a particular equal-time surface is chosen. For example, the proper-time
gauge favors recently thermalized regions, as compared with the scale-factor
gauge. Probability distributions \emph{can} be defined in a gauge-invariant
manner in some cases, for instance in a class of models where an observable
assumes all possible values within one causally connected thermalized patch~\cite{Vil98,VanVilWin00}.
However, a straightforward gauge-invariant definition of probability distributions
is still lacking in the important case where observables assume different values
in causally disconnected patches (for some approaches, see~\cite{GarVil01}). 

In this paper I shall not consider these gauge-invariant prescriptions but instead
focus on the issues of gauge dependence in the usual FP formalism. This apparently
unavoidable gauge dependence has caused some consternation~\cite{Haw03,GraTur05}
regarding the overall validity of the standard picture of eternal inflation.
According to a criticism put forward in Ref.~\cite{Haw03}, inflating regions
cannot be said to dominate the volume of the universe because the volume is
not a gauge-invariant characteristic. One might then ask whether the entire
picture of eternal inflation is not a gauge artifact or a result of inconsistently
applied statistics. For instance, it was found in Ref.~\cite{GraTur05} that
certain volume-weighted observables behave quite similarly in models with and
without eternal inflation, thus casting doubt on the usually claimed ability
of eternal inflation to forget initial conditions. 

The purpose of this paper is to answer these criticisms and to describe the
defining features of eternal inflation more accurately. As I will discuss below,
the mentioned criticisms correctly point out the flaws in certain often made
statements about eternal inflation. Nevertheless, the standard picture of eternal
inflation remains unchanged. The presence of eternal inflation is manifested
by the permanent increase of the number of causally disconnected non-thermalized
regions. On the other hand, answers to certain other questions, especially those
involving the proper 3-volume, turn out to be heavily influenced by the choice
of equal-time surfaces. I shall attempt to clarify the influence of the gauge
by analyzing various specific statements that might be made about observables
in eternally inflating spacetimes. 

The following questions will be answered:

\begin{itemize}
\item Whether the volume of inflating regions is larger than the volume of thermalized
regions. (Depends on the choice of the time gauge.)
\item More generally, whether a volume-weighted value of an observable has a qualitatively
different behavior when eternal inflation occurs. (Depends on the choice of
the time gauge.)
\item Whether it is possible to describe the manifestations of eternal inflation in
a gauge-invariant manner. In particular, whether the volume of the inflating
domain exhibits an unbounded growth with time regardless of the time gauge.
(Yes, no.)
\item Whether the FP formalism admits a particular time gauge such that the resulting
probability distribution is unbiased. (No.)
\end{itemize}

\section{Eternal inflation in a box}

The global geometry of eternally inflating spacetimes possesses certain counter-intuitive
properties. To render the discussion more visual, let us turn to a drastically
simplified toy model of eternal inflation that nevertheless exhibits all the
qualitative features of interest.

\begin{figure}
\begin{center}\psfrag{inflating}{inflating regions}\psfrag{thermalized}{thermalized regions}\includegraphics[%
  width=3.2in]{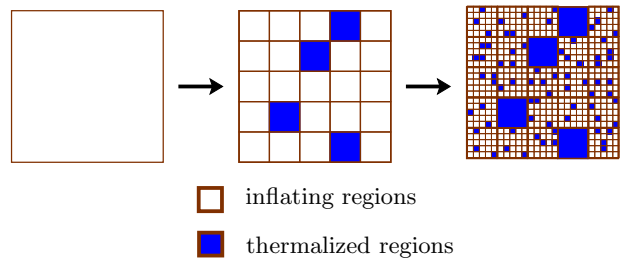}\end{center}

\caption{First steps in the construction of a random Sierpi\'{n}ski carpet with $N=5$
and $q=5/6$. \label{cap:square}}
\end{figure}

In the toy model which we call {}``inflation in a box,'' time elapses in discrete
steps $s=0,1,2,...$ and the space is reduced to two dimensions and further
to the square domain $0<x,y<1$, where $x,y$ are the comoving coordinates measured
in Hubble units $H^{-1}$. The entire initial Hubble-size domain is assumed
to be inflating at time $s=0$. To imitate inflation during one timestep, we
subdivide the initial inflating square into $N\times N$ equal sub-squares of
size $N^{-1}\times N^{-1}$; at the next step, each sub-square will again have
the Hubble proper size. Then we randomly mark some of the smaller squares as
{}``thermalized'' assuming that each Hubble-size inflating square continues
inflation with a probability $q$ (where $0<q<1$) and thermalizes with probability
$1-q$, independently of all other squares. The selection of thermalized squares
concludes the simulation for one timestep. At the next timestep, the same procedure
of subdivision and random thermalization is applied to each Hubble-sized inflating
square, while the {}``thermalized'' squares do not evolve any further (see
Fig.~\ref{cap:square}). This random process was called a {}``random Sierpi\'{n}ski
carpet'' or a {}``curdling'' in the book~\cite{Mandelbrot}.

At a step $s$ we thus have a number of inflating squares with comoving size
$N^{-s}\times N^{-s}$ and also a number of thermalized squares whose comoving
sizes are determined by the times of their thermalization. It is straightforward
to show that a given comoving point $\left(x,y\right)$ remains in the inflating
regime with probability $q^{s}$ after $s$ steps, and that the mean number
of inflating squares at a step $s$ is $\left(N^{2}q\right)^{s}$ while the
mean comoving \emph{}2-volume of the inflating domain is $q^{s}$. Since each
inflating square has Hubble proper size, the number of inflating squares $\left(N^{2}q\right)^{s}$
is naturally interpreted as the \emph{proper} (as opposed to \emph{comoving})
volume of the inflating domain at time $s$. It follows that the mean proper
volume of inflating domain grows without bound if $N^{2}q>1$ and decreases
to zero if $N^{2}q<1$. Thus the condition $N^{2}q>1$ allows future-eternal
inflation to occur with a nonzero probability. (A rigorous derivation of these
results is given in the theory of branching processes, see e.g.~the book~\cite{AthNey72},
chapter 1.) 

This toy model serves as a crude analogy for the process of random nucleation
of bubbles in a de Sitter spacetime (see e.g.~\cite{GW83}). The {}``thermalized
squares'' represent the nucleated bubbles of true vacuum which expand and quickly
reach the comoving Hubble size (computed at nucleation time). The subdivision
of the inflating squares mimics an exponential expansion, while the parameter
$N$ plays the role of the expansion factor during one timestep. Since the thermalized
squares are not further subdivided, the expansion of spacetime within the true
vacuum bubbles is not modeled. As a remedy we consider a variation of the box
model where the thermalized squares remain thermalized but are subdivided into
$M\times M$ parts at each timestep, where $M<N$ is chosen to emulate the slower
expansion of true vacuum regions. This variation of the box model characterized
by the parameters $q,N,M$ will be sufficiently general for the present analysis.

Let us now make a connection of the box model with models of eternal inflation
driven by a scalar field $\phi$ with an effective potential $V(\phi)$. The
FP formalism requires to choose a time slicing parametrized by a time coordinate
$\tau$. A possible choice is $\tau=t$ where $t$ is the proper time measured
along comoving geodesics $\mathbf{x}=\textrm{const}$. Then one considers two
probability distributions defined at a fixed time $\tau$: first, the distribution
$P(\phi,\tau)$ of values along a given (randomly chosen) comoving trajectory,
and second, the total proper volume $P_{V}(\phi,\tau)$ of all regions with
the value $\phi$ at time $\tau$. (Some issues regarding the interpretation
of the distributions $P$ and $P_{V}$ are clarified in Appendix~\ref{sec:Interpretations-of-the}.)
If the time parameter $\tau$ is related to the proper time $t$ by a function
of $\phi$ only, \begin{equation}
d\tau=T(\phi)dt,\quad T(\phi)>0,\label{eq:gen gauge}\end{equation}
 then one can compute the probability distributions $P$ and $P_{V}$ by solving
the FP equations,%
\footnote{Here and below we use the Ito factor ordering in the diffusion terms~\cite{Vil99}.%
}\begin{align}
\partial_{\tau}P & =\partial_{\phi}^{2}\left(D(\phi)P\right)-\partial_{\phi}\left(v(\phi)P\right),\label{eq:FP c}\\
\partial_{\tau}P_{V} & =\partial_{\phi}^{2}\left(D(\phi)P_{V}\right)-\partial_{\phi}\left(v(\phi)P_{V}\right)+3h(\phi)P_{V},\label{eq:FP p}\end{align}
where $D(\phi)$, $v(\phi)$, and $h(\phi)$ are appropriate kinetic coefficients~\cite{WinVil96,Win02}
that depend on the potential $V(\phi)$ and on the function $T(\phi)$. The
precise form of these coefficients will not be used in this paper.

We can describe {}``inflation in a box'' by analogous equations if we assume
that the variable $\phi$ takes on a discrete set of values, e.g.~$\phi=0$
standing for inflation and $\phi=1$ for thermalization. The distributions $P(\phi,\tau)$
and $P_{V}(\phi,\tau)$ are reduced to $P_{j}(s)\equiv P(j,s)$ and $P_{Vj}(s)\equiv P_{V}(j,s)$,
where $j=0,1$ and $s$ is the timestep. The equations describing these distributions
are\begin{align}
P_{0}(s+1) & =qP_{0}(s),\\
P_{1}(s+1) & =P_{1}(s)+\left(1-q\right)P_{0}(s),\\
P_{V0}(s+1) & =N^{2}qP_{V0}(s),\\
P_{V1}(s+1) & =M^{2}P_{V1}(s)+N^{2}\left(1-q\right)P_{V0}(s),\end{align}
where $q,N,M$ are the parameters of the box model. With the initial conditions
$P_{0}(0)=P_{V0}(0)=1$, $P_{1}(0)=P_{V1}(0)=0$, the solutions are\begin{align}
P_{0}(s)=q^{s},\; & P_{V0}(s)=\left(N^{2}q\right)^{s},\quad P_{1}(s)=1-q^{s},\\
P_{V1}(s)= & \frac{1-q}{q}\left(N^{2}q\right)^{s}\frac{1-\lambda^{s}}{1-\lambda},\quad\lambda\equiv\frac{M^{2}}{N^{2}q}.\end{align}
It is clear that the proper volume of the inflating domain $P_{V0}(s)$ grows
without bound if $N^{2}q>1$. This is the usual sign of eternal inflation.

It will be more convenient to analyze the box model in continuous time $t$
rather than in the discrete time $s$. Denoting by $\alpha$ the thermalization
rate and by $H_{0}$, $H_{1}$ the expansion rates of the inflating and the
thermalized domains respectively, we can write the following equations describing
the distributions of the comoving and the proper volume,\begin{align}
\frac{d}{dt}P_{0} & =-\alpha P_{0},\quad\frac{d}{dt}P_{1}=\alpha P_{0},\label{eq:P0 equ}\\
\frac{d}{dt}P_{V0} & =2H_{0}P_{V0}-\alpha P_{V0},\\
\frac{d}{dt}P_{V1} & =2H_{1}P_{V1}+\alpha P_{V0}.\label{eq:Pv equ}\end{align}
With the same initial conditions as above, the solution is\begin{align}
P_{0}(t)=e^{-\alpha t},\; & P_{V0}(t)=e^{(2H_{0}-\alpha)t},\label{eq:PV0 ans}\\
P_{1}(t)=1-e^{-\alpha t},\; & P_{V1}(t)=\frac{e^{(2H_{0}-\alpha)t}-e^{2H_{1}t}}{2(H_{0}-H_{1})-\alpha}\alpha.\label{eq:PV1 ans}\end{align}
The correspondence with the discrete-time model is obtained by setting $q=e^{-\alpha}$
and $H_{0}=\ln N$, $H_{1}=\ln M$. Eternal inflation is possible if $\alpha<2H_{0}$.

An example of a different choice of the time variable is the {}``$e$-folding
time'' $\tau=\ln a$, where $a$ is the scale factor. To derive the analogs
of Eqs.~(\ref{eq:P0 equ})-(\ref{eq:Pv equ}) for the volume distributions
at constant $\tau$, we note that an infinitesimal interval $d\tau$ corresponds
to the proper time interval $dt=H_{0}^{-1}d\tau$ within the inflating domains,
and thus a fraction $\alpha dt=\alpha H_{0}^{-1}d\tau$ of the total inflating
volume will thermalize during the interval $d\tau$. This applies to the comoving
as well as to the proper volume since the inflating domain is homogeneous. Therefore
the increase $dP_{1}$ in the comoving volume of the thermalized domain during
the $e$-folding time interval $d\tau$ will be $\alpha H_{0}^{-1}P_{0}d\tau$.
In this way we arrive to the equations\begin{align}
\frac{d}{d\tau}P_{0} & =-\frac{\alpha}{H_{0}}P_{0},\quad\frac{d}{d\tau}P_{1}=\frac{\alpha}{H_{0}}P_{0},\label{eq:Pt equ}\\
\frac{d}{d\tau}P_{V0} & =2P_{V0}-\frac{\alpha}{H_{0}}P_{V0},\\
\frac{d}{d\tau}P_{V1} & =2P_{V1}+\frac{\alpha}{H_{0}}P_{V0}.\label{eq:Pt equ 2}\end{align}
Note that these equations cannot be obtained from Eqs.~(\ref{eq:P0 equ})-(\ref{eq:Pv equ})
by a simple change of the time variable. The solutions of Eqs.~(\ref{eq:Pt equ})-(\ref{eq:Pt equ 2})
are\begin{align}
P_{0}(\tau)=e^{-\alpha\tau/H_{0}}, & \; P_{V0}(\tau)=e^{(2-\alpha/H_{0})\tau},\\
P_{1}(\tau)=1-e^{-\alpha\tau/H_{0}}, & \; P_{V1}(\tau)=e^{2\tau}-e^{(2-\alpha/H_{0})\tau}.\end{align}

The volume of the inflating domain grows proportionally to $a^{2-\alpha/H_{0}}$
in both the time gauges because the inflationary expansion rate $H_{0}$ is
everywhere the same. A two-dimensional domain whose proper 2-volume grows slower
than $a^{2}$, namely as $a^{\gamma}$ with $\gamma<2$, can be visualized as
an expanding lacunary fractal with dimension $\gamma$, in the following sense.
A lacunary fractal set $S$ having a fractal dimension $\gamma<2$ contains
infinitely many {}``holes'' of diminishing size, and the 2-volume of $S$
vanishes. To obtain a domain with a nonvanishing volume, one can coarse-grain
the set $S$ on a fixed scale which is the Hubble scale in the context of inflation.
We may denote the result of the coarse-graining by $\tilde{S}$; the set $\tilde{S}$
is interpreted as the inflating domain and the holes in $S$ are regions that
will eventually thermalize. At a given time, holes in $S$ having sizes below
the Hubble scale are invisible in the coarse-grained set $\tilde{S}$. As the
set $S$ expands (while the coarse-graining scale is kept fixed), it appears
that new holes are constantly created within the domain $\tilde{S}$, making
its volume grow slower than $a^{2}$. The resulting fractal structure of the
inflating domain was first explored in Ref.~\cite{VilAry89} where it was shown
that a domain expanding as $\propto a^{\gamma}$ is characterized by the fractal
dimension $\gamma$. Thus the fractal dimension of the inflating domain in the
box model is $\gamma=2-\alpha H_{0}^{-1}$. A more rigorous definition of the
fractal dimension of the inflating domain, together with a proof of its gauge
independence, was given in Ref.~\cite{Win02}.

\section{\label{sec:Volume-of-inflating}Volume of inflating vs. thermalized domains}

Let us now consider the question of whether the inflating domain dominates the
volume of the universe at late times. In the box model with continuous time,
the ratio of the total thermalized volume to the total inflating volume at a
proper time $t$ is found from Eqs.~(\ref{eq:PV0 ans})-(\ref{eq:PV1 ans}),\begin{equation}
\left.\frac{V_{\textrm{therm}}}{V_{\textrm{infl}}}\right|_{s}=\frac{P_{V1}(t)}{P_{V0}(t)}=\alpha\frac{1-e^{-\mu t}}{\mu},\quad\mu\equiv2(H_{0}-H_{1})-\alpha.\label{eq:mu def}\end{equation}
At late times $t\rightarrow\infty$, this ratio tends to infinity if $\mu\leq0$
and to a finite nonzero limit $\alpha/\mu$ if $\mu>0$. The value of the latter
limit may be any real number depending on the values of the parameters $H_{0},H_{1},\alpha$.
With the typical assumptions $\alpha\ll2H_{0}$ and $H_{1}<H_{0}$, we find
a finite and small value of the limit, $\alpha/\mu\ll1$, indicating that most
of the volume is in the inflating domain.

However, if we compute the volume ratio at a fixed scale factor rather than
at a fixed step, we obtain a different result,\begin{align}
\left.\frac{V_{\textrm{therm}}}{V_{\textrm{infl}}}\right|_{\tau}=\frac{P_{V1}(\tau)}{P_{V0}(\tau)}= & e^{\alpha\tau/H_{0}}-1.\end{align}
This ratio tends to infinity as $\tau\rightarrow\infty$, indicating that most
of the volume is contained in the thermalized domain, in a direct opposition
to the typical result obtained in the proper time gauge. Therefore the statement
that \emph{the volume is dominated by the inflating domain during eternal inflation}
does not hold in all gauges. Also, the ratio of the inflating to the thermalized
volume does not necessarily characterize the presence or the absence of eternal
inflation.

\section{Volume-weighted observables}

We now consider volume-weighted observables in an eternally inflating spacetime.
A \textsl{volume-weighted} observable is defined using the volume $V(A)$ occupied
by regions that have a certain value $A$ of the observable at a time $t$:\begin{equation}
\left\langle A\right\rangle _{V}=\frac{\sum_{A}AV(A)}{\sum_{A}V(A)}.\end{equation}
 Volume-weighting is always performed along a particular equal-time surface. 

In the box model, the possible candidates for volume weighting are the field
$\phi$ and the scale factor $a$. Since the field $\phi$ serves only to indicate
the inflating or the thermalized state of a region, a volume-weighted average
of $\phi$ or, more generally, of a function $A(\phi)$ at a time $t$ is completely
determined by the distributions $P_{V0}(t)$ and $P_{V1}(t)$, i.e.\begin{equation}
\left\langle A(\phi)\right\rangle _{V}=\frac{A(0)P_{V0}+A(1)P_{V1}}{P_{V0}+P_{V1}}=\frac{A(0)+A(1)\frac{P_{V1}}{P_{V0}}}{1+\frac{P_{V1}}{P_{V0}}}.\end{equation}
 As we have seen, the ratio $P_{V1}/P_{V0}$ is highly gauge-dependent and so
are the volume-weighted averages of $\phi$ or of any functions of $\phi$.

Turning now to the scale factor $a$, we first note that in the scale factor
gauge the entire equal-time hypersurface has the same value of $a$ and it is
useless to perform a volume-weighted averaging. Therefore we consider the proper
time gauge, where the volume-weighted average of the scale factor differs from
its statistical mean. To describe the distribution of the scale factor observed
at a proper time $t$ at a given comoving point in the box model, it suffices
to compute the probability density $p_{th}(s)$ for the given point to thermalize
within a time interval $[s,s+ds]$, \begin{equation}
p_{th}(s)ds=e^{-\alpha s}\alpha ds.\end{equation}
It is clear that $p_{th}(s)$ is the also the probability density for the scale
factor to assume the value $a=e^{H_{0}s+H_{1}(t-s)}$ at time $t\geq s$ at
a randomly chosen comoving point. It is more convenient to perform computations
with the number of $e$-foldings, $\ln a$, for which we have the statistical
mean\begin{equation}
\left\langle \ln a\right\rangle =\left\langle sH_{0}+\left(t-s\right)H_{1}\right\rangle =H_{1}t+(H_{0}-H_{1})\left\langle s\right\rangle \end{equation}
and the volume-weighted mean\begin{equation}
\left\langle \ln a\right\rangle _{V}=H_{1}t+(H_{0}-H_{1})\left\langle s\right\rangle _{V},\end{equation}
so it remains to compute the means $\left\langle s\right\rangle $ and $\left\langle s\right\rangle _{V}$
using the distribution $p_{th}(s)$. Calculations yield\begin{equation}
\left\langle s\right\rangle =\int_{0}^{t}p_{th}(s)sds=\frac{1-e^{-\alpha t}}{\alpha}-te^{-\alpha t}\end{equation}
and\begin{align}
\left\langle s\right\rangle _{V} & =\frac{\int_{0}^{t}p_{th}(s)e^{2H_{0}s+2H_{1}(t-s)}sds}{\int_{0}^{t}p_{th}(s)e^{2H_{0}s+2H_{1}(t-s)}ds}=\frac{\mu te^{\mu t}-e^{\mu t}+1}{\left(e^{\mu t}-1\right)\mu},\end{align}
where $\mu$ is defined in Eq.~(\ref{eq:mu def}). Depending on the value of
the parameter $\mu$, the volume-weighted average for late times $t$ may be
dominated by the contribution of either inflating or thermalized regions. For
instance, if $\mu\leq0$ (domination by thermalized regions) then \begin{equation}
\lim_{t\rightarrow\infty}\left\langle s\right\rangle _{V}=-\frac{1}{\mu}\end{equation}
 and thus for large $t$ we have\begin{equation}
\left\langle \ln a\right\rangle _{V}=H_{1}t-\frac{1}{\mu}(H_{0}-H_{1})\approx H_{1}t,\end{equation}
which is almost the same as the statistical mean, \begin{equation}
\left\langle \ln a\right\rangle =H_{1}t+\frac{1}{\alpha}(H_{0}-H_{1}).\end{equation}
On the other hand, if $\mu>0$ (domination by inflating regions) then $\left\langle s\right\rangle _{V}\approx t$,
i.e.~almost all volume is filled by very recently thermalized regions, and
so\begin{equation}
\left\langle \ln a\right\rangle _{V}\approx H_{1}t+(H_{0}-H_{1})t=H_{0}t.\end{equation}
Note that the possibility of eternal inflation is compatible with both cases
$\mu>0$ and $\mu\leq0$. We conclude that volume-weighted observables such
as $\left\langle \ln a\right\rangle _{V}$ generally furnish neither any information
specific to eternal inflation nor any gauge-invariant information.

\section{Growth of volume of inflating domains}

The hallmark of eternal inflation is the unbounded increase in the total number
of independent inflating regions. It is also often stated that the total proper
3-volume of the inflating domain grows without bound at late times in an eternally
inflating spacetime; this is certainly true for the 3-volume computed along
hypersurfaces of equal proper time or of equal scale factor. One of the criticisms
expressed in Refs.~\cite{Haw03,GraTur05} was that the 3-volume is a gauge-dependent
characteristic and thus cannot be used as an indication of the presence or the
absence of eternal inflation. In an earlier attempt to resolve this issue, it
was proven in Ref.~\cite{Win02} (Sec.~III A) that an eternally inflating
spacetime manifests an unbounded volume growth in all time gauges of the form~(\ref{eq:gen gauge}).
However, one may imagine time gauges not of the form~(\ref{eq:gen gauge}).
Equal-time probability distributions in such gauges cannot be described by the
FP equations~(\ref{eq:FP c})-(\ref{eq:FP p}) where the kinetic coefficients
depend only on the field $\phi$. 

In fact, the statement about the growing volume of the inflating domain \emph{is}
gauge-dependent. If one allows completely arbitrary time slicings, the total
volume of the inflating domain during eternal inflation might either grow or
decrease depending on the time slicing. I illustrate this dramatic dependence
on the choice of gauge in Appendix~\ref{sec:Counterexamples-to-the} where
I present an explicitly constructed foliation of a de Sitter spacetime by spacelike
hypersurfaces whose proper 3-volume monotonically decreases to zero at late
times. These hypersurfaces are artificially chosen to be almost null almost
everywhere, which makes their 3-volume arbitrarily small despite the expansion
of the background spacetime. Since an inhomogeneous eternally inflating universe
expands {}``slower'' than a pure de Sitter universe, the same family of hypersurfaces
will also have a decreasing 3-volume in inhomogeneous inflating universes. It
follows that the 3-volume of an \emph{arbitrary} family of equal-time hypersurfaces
cannot be used as a criterion for the presence of eternal inflation.

However, a weaker statement is sufficient: namely, eternal inflation is present
if (and only if) there \emph{exists} a choice of time slicing with an unbounded
growth of the 3-volume of inflating domains. In the remainder of this section
we shall demonstrate the equivalence of this criterion and the usual definition
of eternal inflation (the existence of inflating domains at arbitrary late times).
Note that the existence of something {}``at an arbitrarily late time'' is
a gauge-independent notion due to the monotonicity of time parameters in all
gauges.

If the 3-volume of the inflating domain grows without bound in some time gauge,
then obviously there exist inflating domains at arbitrarily late times. It remains
to prove that if inflating domains exist arbitrarily late, then the inflating
volume grows in \emph{some} gauge. This growth can be most easily demonstrated
in the $e$-folding gauge, $\tau=\ln a$, using the construction of {}``eternal
comoving points''~\cite{Win02}. These points follow comoving geodesic worldlines
that forever remain within the inflating domain and never enter any thermalized
regions. It was shown in Ref.~\cite{Win02} using general topological arguments
that the presence of inflating domains at arbitrarily late times entails the
existence of infinitely many such {}``eternal points.'' Since each eternal
point is surrounded by at least a Hubble-sized inflating region at any time,
the 3-volume of the inflating domain at $e$-folding time $\tau$ is greater
than $H^{-3}N$, where $N$ is the number of eternal points that are pairwise
separated by a proper distance of at least $H^{-1}$ at time $\tau$. To show
that the volume is greater than a preassigned bound $M$, we thus need to find
at least $N_{0}=MH^{3}$ widely separated eternal points at a sufficiently late
time. Since the eternal points are infinitely numerous, we can always choose
a subset $S$ containing $N_{0}$ such points and then determine the minimum
comoving distance $\delta x$ between any two points from the subset $S$. Within
a hypersurface of constant scale factor $a=e^{\tau}$, the proper Hubble distance
$H^{-1}$ corresponds to the comoving distance $H^{-1}e^{-\tau}$. For sufficiently
late $e$-folding times $\tau>\tau_{0}\equiv-\ln(H\delta x)$ the scale factor
$a=e^{\tau}$ is large enough so that the proper distance between any two points
from $S$ is larger than $H^{-1}$. Thus for $\tau>\tau_{0}$ there exist at
least $N_{0}$ independent inflating regions, each having a volume at least
$H^{-3}$. So we have shown that the volume of the inflating domain for $\tau>\tau_{0}$
is greater than $N_{0}H^{-3}=M$. Since $M$ is an arbitrary bound, we conclude
that the 3-volume computed in the $e$-folding gauge grows without bound. It
follows that an unbounded growth of the 3-volume is found also in all gauges
of the form~(\ref{eq:gen gauge}), including the proper time gauge.

\section{Is there a {}``correct'' time parameter?}

As we have seen in Sec.~\ref{sec:Volume-of-inflating}, the division of 3-volume
between the inflating and the thermalized domains is strongly gauge-dependent.
For instance, the proper-time distribution is biased towards regions with faster
expansion rates, while the scale factor gauge favors regions with larger comoving
volumes. The probability distribution $P(\phi,\tau)$ necessarily depends on
the choice of the time parameter $\tau$. It has been sometimes proposed that
a certain choice of the parameter $\tau$ will yield physically justified, unbiased
probability distributions. In this section I shall demonstrate that no choice
of the time parameter guarantees unbiased results, at least in a certain class
of inflationary models.

\begin{figure}
\begin{center}\psfrag{phi}{$\phi$}\psfrag{V}{$V$}

\psfrag{ps2}{$\phi_{*}^{(1)}$}\psfrag{ps1}{$\phi_{*}^{(2)}$}

\psfrag{p2}{$\phi_1$}\psfrag{p1}{$\phi_2$}\includegraphics[%
  width=3.2in]{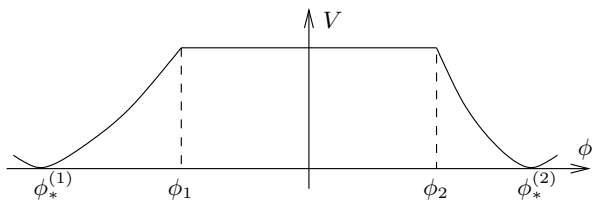}\end{center}

\caption{Illustrative inflationary potential with a flat self-reproduction regime $\phi_{1}<\phi<\phi_{2}$
and deterministic regimes $\phi_{*}^{(1)}<\phi<\phi_{1}$ and $\phi_{2}<\phi<\phi_{*}^{(2)}$.
\label{cap:pot1}}
\end{figure}

In the relevant class of models, the self-reproduction regime is symmetric with
respect to the observable parameters; a specific model of this kind was considered
in Ref.~\cite{Vil98}. For the present purposes it is sufficient to analyze
a toy model of scalar-field inflation with an effective potential shown in Fig.~\ref{cap:pot1}.
Our analysis shall partially follow Ref.~\cite{Vil95} where a similar potential
was introduced. The potential is flat in the range $\phi_{1}<\phi<\phi_{2}$
where the evolution is fluctuation-dominated, while the evolution of regions
with $\phi>\phi_{2}$ or $\phi<\phi_{1}$ is completely deterministic (fluctuation-free).
It is assumed that the fluctuation-dominated range $\phi_{1}<\phi<\phi_{2}$
is sufficiently wide to cause an eternal self-reproduction of inflating regions.
There are two thermalization points, $\phi=\phi_{*}^{(1)}$ and $\phi=\phi_{*}^{(2)}$,
which may be associated to different types of true vacuum and thus to different
observed values of cosmological parameters. The question is to compare the volumes
${\cal V}_{1,2}$ of regions thermalized into these two vacua. Since there is
an infinite volume thermalized into either vacuum, one hopes to obtain a sensible
answer for the volume ratio ${\cal V}_{1}/{\cal V}_{2}$.

In the present case, due to the symmetry of the potential it is natural to assume
that Hubble-sized regions exiting the eternal self-reproduction regime at $\phi=\phi_{1}$
and at $\phi=\phi_{2}$ are equally abundant. Since the evolution within the
ranges $\phi_{*}^{(1)}<\phi<\phi_{1}$ and $\phi_{2}<\phi<\phi_{*}^{(2)}$ is
deterministic, the regions exiting the self-reproduction regime at $\phi=\phi_{1}$
or $\phi=\phi_{2}$ will experience a fixed number of $e$-foldings which we
may denote $\ln Z_{1}$ and $\ln Z_{2}$ respectively. Therefore the volume
of regions thermalized at $\phi=\phi_{*}^{(j)}$, where $j=1,2$, will be increased
by the factors $Z_{j}^{3}$ and thus the volume ratio is\begin{equation}
\frac{{\cal V}_{1}}{{\cal V}_{2}}=\frac{Z_{1}^{3}}{Z_{2}^{3}}.\label{eq:ans2}\end{equation}

We shall now compare this result with that of the gauge-dependent FP approach.
The time-dependent distribution of the 3-volume $P_{V}(\phi,\tau)$ is a solution
of the FP equation~(\ref{eq:FP p}) in a time gauge $\tau$. The late-time
behavior of the volume distribution is\begin{equation}
P_{V}(\phi,\tau)=P_{V}^{(0)}(\phi)e^{\gamma\tau},\label{eq:Pv0}\end{equation}
where $\gamma>0$ is the largest eigenvalue of the corresponding stationary
FP equation. (It was shown in Ref.~\cite{Win02} that $\gamma>0$ in all gauges
$\tau$ admitted by the FP formalism.) Since the volume of regions with all
values of $\phi$ grows without bound, one needs to introduce a cutoff to compare
the total volumes ${\cal V}_{1,2}$ thermalized at $\phi=\phi_{*}^{(1)}$ and
$\phi=\phi_{*}^{(2)}$. The \textsl{equal-time cutoff}, i.e.~the procedure
where one counts only the volume of the regions thermalized before a fixed time
$\tau=\tau_{\max}$, is implemented straightforwardly in the FP formalism and
in the present case yields~\cite{Vil95} \begin{equation}
\frac{{\cal V}_{1}}{{\cal V}_{2}}=\frac{Z_{1}^{3}}{Z_{2}^{3}}\exp\left[-\gamma(\Delta\tau_{1}-\Delta\tau_{2})\right],\label{eq:ans1}\end{equation}
where $\Delta\tau_{j}$, $j=1,2$, are the time intervals elapsed in the time
gauge $\tau$ during the deterministic evolution of the field from $\phi_{j}$
to $\phi_{*}^{(j)}$. 

One can derive Eq.~(\ref{eq:ans1}) from elementary considerations as follows.
A Hubble region exiting the self-reproduction regime at $\phi=\phi_{1}$ will
thermalize after a fixed duration of time $\Delta\tau_{1}$ and will accumulate
the growth factor $Z_{1}$ during this time. Similarly, regions exiting at $\phi=\phi_{2}$
will reach the thermalization point $\phi=\phi_{*}^{(2)}$ after a time interval
$\Delta\tau_{2}$ and accumulate the growth factor $Z_{2}$. The {}``arrival
times'' $\Delta\tau_{j}$ and the growth factors $Z_{j}$, $j=1,2$, can be
computed using the particular shape of the potential $V(\phi)$ but we shall
not need their explicit forms.

It follows that the regions thermalized at $\phi=\phi_{*}^{(j)}$ at time $\tau$
have previously exited the self-reproduction regime at time $\tau-\Delta\tau_{j}$.
According to Eq.~(\ref{eq:Pv0}), the volume of regions exiting at $\phi=\phi_{j}$
at time $\tau-\Delta\tau_{j}$ is\begin{equation}
P_{V}^{(0)}(\phi_{j})\exp\left[\gamma(\tau-\Delta\tau_{j})\right],\end{equation}
 where $P_{V}^{(0)}(\phi_{1})=P_{V}^{(0)}(\phi_{2})\equiv P_{V}^{(0)}$ due
to the symmetry of the potential in the self-reproduction regime. The total
volume of regions thermalized at $\phi=\phi_{*}^{(j)}$ up to a time $\tau_{\max}$
is then \begin{align}
{\cal V}_{j}(\tau_{\max}) & =P_{V}^{(0)}Z_{j}^{3}\int_{0}^{\tau_{\max}}\exp\left[\gamma(\tau-\Delta\tau_{j})\right]\nonumber \\
 & \propto Z_{j}^{3}\exp\left[\gamma(\tau_{\max}-\Delta\tau_{j})\right],\end{align}
where we have omitted $j$-independent factors. Thus we obtain the volume ratio
as the limit \begin{align}
\frac{{\cal V}_{1}}{{\cal V}_{2}} & =\lim_{\tau_{\max}\rightarrow\infty}\frac{{\cal V}_{1}(\tau_{\max})}{{\cal V}_{2}(\tau_{\max})}=\frac{Z_{1}^{3}}{Z_{2}^{3}}\exp\left[-\gamma(\Delta\tau_{1}-\Delta\tau_{2})\right].\end{align}
This coincides with Eq.~(\ref{eq:ans1}).

As already noted in Ref.~\cite{Vil95}, the reason for the gauge dependence
of the volume ratio~(\ref{eq:ans1}) is the presence of the times $\Delta\tau_{j}$
which generally leads to an appreciable bias. The arrival times $\Delta\tau_{j}$
are generally different, $\Delta\tau_{1}\neq\Delta\tau_{2}$, unless the inflaton
potential $V(\phi)$ is completely symmetric.%
\footnote{The arrival times would be equal if it were possible to use the hypersurfaces
of equal $\phi$ as the time slicing. However, a hypersurface $\phi=\textrm{const}$
is in general not spacelike because the evolution of $\phi$ is not monotonic
in the self-reproduction regime.%
} Since $\gamma>0$ in all gauges, it is impossible to choose a gauge $\tau$
in which the results~(\ref{eq:ans2}) and (\ref{eq:ans1}) would coincide for
all potentials. We conclude that there exists no {}``correct'' gauge $\tau$
that would always guarantee unbiased results for the equal-time cutoff procedure.

\section*{Acknowledgments}

The author is grateful to Matthew Parry for stimulating conversations and to
Alex Vilenkin for discussions and comments on the manuscript. Thanks are also
due to the referee who made some useful suggestions.

\appendix

\section{\label{sec:Interpretations-of-the}Interpretations of the Fokker-Planck formalism}

In this appendix I shall clarify the possible interpretations of the distributions
$P(\phi,\tau)$ and $P_{V}(\phi,\tau)$ in terms of an ensemble of imaginary
comoving observers placed throughout the inflationary spacetime.

The comoving volume distribution $P(\phi,\tau)$ can be found as a solution
of the FP equation~(\ref{eq:FP c}) as well as via a Langevin equation. The
Langevin and the Fokker-Planck equations are equivalent descriptions of the
same stochastic process, and we shall focus on the FP description. 

The distribution $P(\phi,\tau)$ allows two interpretations. On the one hand,
$P(\phi,\tau)$ is the probability for the field to have a value $\phi$ at
time $\tau$ along a single, randomly chosen comoving trajectory; on the other
hand, $P(\phi,\tau)$ is the fraction of comoving volume where the field has
the value $\phi$, within an equal-$\tau$ hypersurface. (The {}``comoving
3-volume'' of a subdomain $S$ of an equal-$\tau$ hypersurface is defined
as the 3-volume of the subdomain $S_{0}$ of the hypersurface $\tau=0$, where
$S_{0}$ consists of points whose comoving worldlines pass through $S$ at time
$\tau$.)

The equivalence of these interpretations can be seen from the following argument.
Consider a large but finite domain of volume $V_{0}$ within the hypersurface
$\tau=0$ and a very dense grid of comoving trajectories starting out from this
domain. The total number $N$ of these trajectories may be chosen as finite
but very large, and the initial volume $V_{0}$ can be subdivided into $N$
tiny regions of comoving volume $V_{0}/N$ situated around each of the $N$
comoving lines. If the initial volume $V_{0}$ is sufficiently large so that
all the statistically possible histories are sufficiently well sampled, then
the number of lines that reach a value $\phi$ at a time $\tau$ is very close
to $NP(\phi,\tau)$. Thus the comoving volume of the subdomain of the equal-$\tau$
hypersurface having the value $\phi$ is $V_{0}P(\phi,\tau)$. Therefore in
the statistical limit $N\rightarrow\infty$ and $V_{0}\rightarrow\infty$, the
fraction of the comoving volume containing the field value $\phi$ at the time
$\tau$ is equal to $P(\phi,\tau)$.

Conversely, if $P(\phi,\tau)$ is the statistical distribution of the comoving
volume along an equal-$\tau$ hypersurface, then the comoving volume of the
subdomain with the value $\phi$ will be $V_{0}P(\phi,\tau)$. We denote this
subdomain by $S(\phi,\tau)$ and consider a comoving trajectory randomly chosen
from the total comoving volume $V_{0}$. The probability of this trajectory
passing through the subdomain $S(\phi,\tau)$ is equal to $V_{0}P(\phi,\tau)/V_{0}=P(\phi,\tau)$.
Therefore $P(\phi,\tau)$ is the probability of the field having the value $\phi$
at time $\tau$ along a randomly chosen trajectory.

We note that there do exist correlations between the values of $\phi$ at nearby
comoving trajectories. However, these correlations are not reflected by the
distribution $P(\phi,\tau)$ since it specifies only the total volume, but not
the location, of regions with the field value $\phi$. If the initial comoving
volume $V_{0}$ is sufficiently large (many Hubble volumes), the distribution
of the field $\phi$ at time $\tau$ along a randomly chosen comoving trajectory
will be $P(\phi,\tau)$ despite the correlations between nearby trajectories.

Turning now to the distribution of the proper volume $P_{V}(\phi,\tau)$, we
note that there is now only one interpretation, namely $P_{V}(\phi,\tau)$ is
the total proper 3-volume of the subdomain with the field value $\phi$ within
an equal-$\tau$ hypersurface. (As before, we consider only the comoving future
of a large but finite initial volume $V_{0}$ at time $\tau=0$.) It appears
to be impossible to interpret $P_{V}(\phi,\tau)$ directly in terms of observations
made on a single randomly chosen comoving trajectory. The distribution $P_{V}(\phi,\tau)$
describes a branching diffusion process with a possibly unlimited growth, and
its interpretation necessarily has to involve a potentially very large ensemble
of trajectories.

A connection can be formulated between the distribution $P_{V}(\phi,\tau)$
and the joint comoving distribution of the field and the scale factor $P(\phi,a,\tau)$.
The comoving distribution $P(\phi,a,\tau)$ is interpreted as the fraction of
the comoving volume where the field has the value $\phi$ and the scale factor
has the value $a$, within an equal-$\tau$ hypersurface. Alternatively, $P(\phi,a,\tau)$
is the probability of the field having the value $\phi$ and the scale factor
having the value $a$ at time $\tau$ along a single, randomly chosen comoving
trajectory. The distribution $P(\phi,a,\tau)$ is the solution of the FP equation\begin{equation}
\partial_{\tau}P=\hat{L}_{\phi}P-\partial_{a}\left(h(\phi)aP\right),\end{equation}
where \begin{equation}
\hat{L}_{\phi}P\equiv\partial_{\phi}^{2}\left(D(\phi)P\right)-\partial_{\phi}\left(v(\phi)P\right)\end{equation}
 is the differential operator entering Eq.~(\ref{eq:FP c}) and $h(\phi)$
is the {}``expansion rate'' function defined by\begin{equation}
\frac{da}{d\tau}=h(\phi)a.\end{equation}
For example, in the proper-time gauge $h(\phi)=H(\phi)$ while in the $e$-folding
gauge $h\equiv1$. 

To obtain the distribution $P_{V}(\phi,\tau)$, one can use the procedure of
{}``volume-weighting,'' \begin{equation}
P_{V}(\phi,\tau)\equiv\int_{0}^{\infty}da\, a^{3}P(\phi,a,\tau),\end{equation}
which takes into account the expansion factor $a^{3}$ along the trajectory.
It is easy to verify that the resulting (unnormalized) distribution $P_{V}(\phi,\tau)$
satisfies the FP equation~(\ref{eq:FP p}). This establishes a link between
the volume interpretation of the distribution $P_{V}(\phi,\tau)$ and properties
of a single comoving trajectory embodied by the joint probability $P(\phi,a,\tau)$.

\section{\label{sec:Counterexamples-to-the}Counterexamples to the growth of volume}

In this section I shall present two examples of foliations described by spacelike
hypersurfaces \begin{equation}
t=t(\tau,\mathbf{x}),\quad-\infty<\tau<+\infty,\end{equation}
 such that the \emph{total} proper 3-volume of each equal-$\tau$ hypersurface
is finite and monotonically \emph{decreases} with $\tau$, asymptotically approaching
zero at late times. The two examples will be foliations of the Minkowski spacetime
and of the future part of a de Sitter spacetime, the latter obviously interpreted
as a homogeneous, eternally inflating universe. These examples unequivocally
demonstrate that an unbounded growth of the 3-volume of equal-time hypersurfaces
during eternal inflation is not observed in all gauges.

\begin{figure}
\begin{center}\psfrag{L}{$L$}\psfrag{2L}{$2L$}\psfrag{3L}{$3L$}\psfrag{-L}{$-L$}\psfrag{-2L}{$-2L$}

\psfrag{S0}{$t$}\psfrag{0}{$0$}\psfrag{x}{$x$}\includegraphics[%
  width=2.5in]{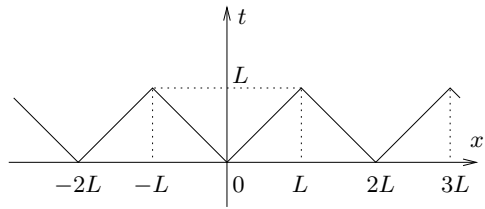}\end{center}

\caption{The piecewise-null surface $S_{0}(x)$ having zero proper length. \label{cap:s0}}
\end{figure}

I shall first describe the idea behind the construction of the foliations and
later give technical details. For simplicity, let us begin by considering a
flat 1+1-dimensional spacetime with standard coordinates $\left(t,x\right)$
and a saw-shaped hypersurface\begin{equation}
t=S_{0}\left(x\right)\equiv\frac{L}{\pi}\cos^{-1}\left(\cos\frac{\pi x}{L}\right),\label{eq:S0 def}\end{equation}
where the inverse cosine function is by definition such that $0\leq S_{0}(x)\leq L$.
The hypersurface $t=S_{0}(x)$ is piecewise null and has kinks at points $x=nL$,
$n\in\mathbb{Z}$ (see Fig.~\ref{cap:s0}). It is clear that the proper length
(i.e.~the {}``1-volume'') of this hypersurface is equal to zero. We can then
choose a smooth spacelike hypersurface situated sufficiently close to $t=S_{0}(x)$,
so that the proper length of the portion of that hypersurface corresponding
to $\left|x\right|<L$ is smaller than a chosen bound $\varepsilon$. It is
also possible to find another hypersurface for $L<|x|<2L$ approaching the null
surface $S_{0}(x)$ even closer, so that the corresponding proper length is
smaller than $\varepsilon/2$, and so on. Thus there exists an \emph{unbounded}
spacelike hypersurface $S_{\varepsilon}(x)$ with a \emph{finite} and arbitrarily
small proper length ${\cal L}[S_{\varepsilon}]<\varepsilon$. 

The Minkowski space can be foliated by copies of the surface $t=S_{0}(x)$ shifted
in the $t$ direction. Instead of this {}``null foliation'' we consider a
family of spacelike hypersurfaces labeled by a time parameter $\tau$,\begin{equation}
t=t\left(\tau,x\right)\equiv\tau+S_{\varepsilon(\tau)}(x),\label{eq:family Mink}\end{equation}
where $\varepsilon(\tau)$ is some function. The family~(\ref{eq:family Mink})
will be an admissible foliation of the 1+1-dimensional Minkowski spacetime as
long as $\partial t/\partial\tau>0$ and each hypersurface is spacelike, \begin{equation}
\left.\left(dt^{2}-dx^{2}\right)\right|_{t=t\left(\tau,x\right),\,\tau=\textrm{const}}<0.\end{equation}
These conditions can be satisfied by a suitable choice of the function $\varepsilon(\tau)$,
and moreover it is possible to have $\varepsilon(\tau)$ monotonically decrease
to zero as $\tau\rightarrow\infty$. The resulting family of hypersurfaces is
the desired foliation of the Minkowski space with a finite and monotonically
decreasing proper volume.

This construction can be generalized to the 3+1-dimensional Minkowski spacetime
by considering a family of spherically symmetric hypersurfaces\begin{equation}
t=t\left(\tau,r\right)=\tau+S_{\varepsilon(\tau)}(r),\quad r\equiv\left|\mathbf{x}\right|.\end{equation}
As will be demonstrated below, the function $\varepsilon(\tau)$ can be chosen
to achieve a finite proper volume ${\cal V}(\tau)$ of the entire equal-$\tau$
hypersurface and moreover to have ${\cal V}(\tau)\rightarrow0$ as $\tau\rightarrow\infty$.

\begin{figure}
\begin{center}\psfrag{L}{$L$}\psfrag{2L}{$2L$}\psfrag{3L}{$3L$}\psfrag{-L}{$-L$}\psfrag{-2L}{$-2L$}

\psfrag{S0}{$\eta$}\psfrag{0}{$0$}\psfrag{x}{$x$}\includegraphics[%
  width=2.5in]{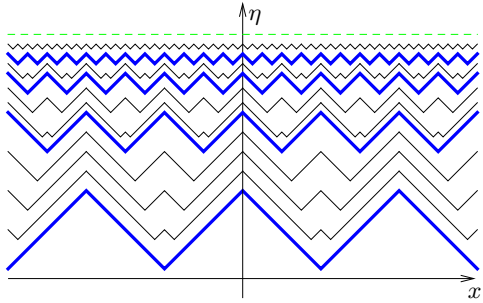}\end{center}

\caption{A null foliation of the de Sitter spacetime in conformal coordinates $\left(\eta,x\right)$.
Thicker lines are the consecutively smaller copies of the saw-shaped surface
$S_{0}(x)$, while the thinner lines in between show the continuous deformation
procedure ({}``growing ridges''). The dashed line corresponds to an infinite
future ($\eta\rightarrow0$). \label{cap:s1}}
\end{figure}

A slightly modified but basically similar construction yields an analogous family
of spacelike hypersurfaces foliating a de Sitter spacetime. For simplicity we
consider a 1+1-dimensional de Sitter spacetime with flat spatial sections. Then
we can use conformal coordinates $\left(\eta,x\right)$, so that the de Sitter
line element is conformally flat,\begin{align}
ds^{2} & =dt^{2}-e^{-2Ht}dx^{2}=\frac{1}{H^{2}\eta^{2}}\left(d\eta^{2}-dx^{2}\right),\\
\eta & =-\frac{1}{H}e^{-Ht},\quad-\infty<\eta<0.\end{align}
All lines $\eta=\eta_{0}\pm x$ are null geodesics and thus the saw-shaped hypersurface
$\eta=S_{0}(x)$ is piecewise null. However, we cannot use the hypersurface
$\eta=S_{0}(x)$ to foliate the spacetime as in the Minkowski case because the
boundary $\eta=0$ prevents us from shifting $S_{0}(x)$ arbitrarily in $\eta$.
Instead it is possible to continuously deform the {}``saw'' and to decrease
the size of its ridges while growing new ridges as time passes (see Fig.~\ref{cap:s1}).
The resulting family of null hypersurfaces $S_{0}(\tau,x)$ can be labeled by
a conformal time parameter $\tau$, where $\tau<0$, $\tau\rightarrow0$ as
$\eta\rightarrow0$, and it is clear from the figure that $S_{0}(\tau,x)$ is
a foliation of the entire future part of the de Sitter spacetime. The piecewise-null
hypersurfaces $S_{0}(\tau,x)$ are then used as a {}``skeleton'' for a family
of non-intersecting, smooth spacelike hypersurfaces $\eta=H_{\varepsilon}(\tau,x)$
with a finite proper volume ${\cal V}[H_{\varepsilon}]<\varepsilon$. It is
possible to choose the function $\varepsilon(\tau)$ such that $\varepsilon(\tau)\rightarrow0$
as $\tau\rightarrow0$. This procedure constructs a foliation of the de Sitter
spacetime with the total proper volume decreasing to zero at late times.

In the remainder of this appendix I describe the required hypersurfaces explicitly
and verify their properties as claimed. 

\begin{figure}
\begin{center}\psfrag{1}{$1$}\psfrag{-1}{$-1$}

\psfrag{C}{$C(x)$}\psfrag{0}{$0$}\psfrag{x}{$x$}\includegraphics[%
  width=2.5in]{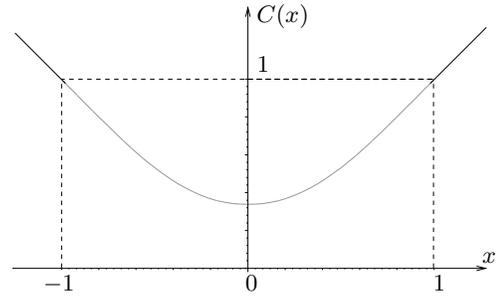}\end{center}

\caption{The ''smooth corner'' function $C(x)$ defined by Eq.~(\ref{eq:c0 def})
with $A=0.01$, $B=0.5$, $C_{1}\approx0.6613$, $C_{2}\approx3.323$. \label{cap:corner0}}
\end{figure}

The first step is to replace the angular piecewise-null hypersurface $S_{0}(x)$
by a smooth hypersurface $S_{0}(x;\varepsilon)$ which is null everywhere except
the {}``caps'' at the corners. To this end we use a {}``smooth corner''
function $C(x)$ which is infinitely differentiable, convex, and satisfies $C(x)=\left|x\right|$
for $\left|x\right|\geq1$ and $\left|C'\right|<1$ for $\left|x\right|<1$.
An example of a suitable function $C(x)$ is plotted in Fig.~\ref{cap:corner0}
and is defined for $\left|x\right|<1$ by\begin{equation}
C(x)\equiv1-C_{1}+C_{2}\int_{0}^{\left|x\right|}dy\int_{0}^{y}dz\,\exp\left(-\frac{A}{z}-\frac{B}{1-z}\right),\label{eq:c0 def}\end{equation}
where $A,B$ are positive constants and\begin{align}
C_{1} & \equiv C_{2}\int_{0}^{1}dy\int_{0}^{y}dz\,\exp\left(-\frac{A}{z}-\frac{B}{1-z}\right),\\
C_{2} & \equiv\left[\int_{0}^{1}dz\,\exp\left(-\frac{A}{z}-\frac{B}{1-z}\right)\right]^{-1}.\end{align}
The function $C(x)$ is smooth because by construction all derivatives of $C(x)$
vanish at $x=0$ and at $x=\pm1$ {[}except $C'(\pm1)=\pm1${]}, while at other
$x$ the function $C(x)$ is analytic. Note also the inequality\begin{equation}
0\leq C(x)-xC'(x)<1\label{eq:C inequ}\end{equation}
which follows from the convexity of $C(x)$.

\begin{figure}
\begin{center}\psfrag{L}{$L$}\psfrag{t}{$t$}\psfrag{-L}{$-L$}\psfrag{Ct}{$\tilde{C}_\delta(x)$}

\psfrag{Cd}{$C_\delta(x)$}\psfrag{d}{$\delta$}\psfrag{-d}{$-\delta$}

\psfrag{C}{$C(x)$}\psfrag{0}{$0$}\psfrag{x}{$x$}\includegraphics[%
  width=2.5in]{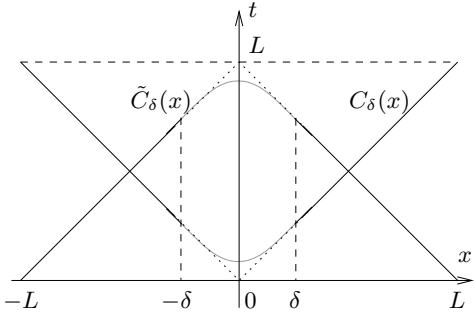}\end{center}

\caption{The rescaled {}``smooth corner'' functions $C_{\delta}(x)$ (corner down)
and $\tilde{C}_{\delta}(x)$ (corner up) defined by Eq.~(\ref{eq:CDdef}).
\label{cap:cornerCD}}
\end{figure}

A rescaled {}``smoothed corner'' function $C(x)$ will be used to replace
small neighborhoods of the top and the bottom corners of the curve $S_{0}(x)$.
Namely, we define the {}``rescaled corner'' functions\begin{equation}
C_{\delta}(x)\equiv C\left(x\delta^{-1}\right)\delta,\qquad\tilde{C}_{\delta}(x)\equiv L-C_{\delta}(x),\label{eq:CDdef}\end{equation}
plotted in Fig.~\ref{cap:cornerCD}. A spherically symmetric hypersurface $t=C_{\delta}(\left|\mathbf{x}\right|)$
is null except for a small range $\left|\mathbf{x}\right|<\delta$, and its
proper 3-volume is smaller than the volume of a sphere with radius $\delta$.
We now replace the corners at $\left|x\right|=nL$ in the curve $S_{0}(x)$
by smoothed corners $C_{\delta_{n}}(x)$ or $\tilde{C}_{\delta_{n}}(x)$, where
$\delta_{n}$ decreases with $n$. For later convenience we set\begin{equation}
\delta_{0}\equiv\varepsilon L;\quad\delta_{n}\equiv\varepsilon Ln^{-2}2^{-n},\quad n=1,2,...,\end{equation}
 where $\varepsilon<1$ is a free dimensionless parameter. To be definite, let
us choose the smoothed-out version of $S_{0}(x)$ as follows,\begin{align}
 & S_{0}(x;\varepsilon)\equiv\nonumber \\
 & \left\{ \begin{array}{l}
C_{\delta_{n}}(x-2nL),\quad2n-\frac{1}{2}\leq\frac{\left|x\right|}{L}\leq2n+\frac{1}{2};\\
\tilde{C}_{\delta_{n}}(x-\left(2n+1\right)L),\quad2n+\frac{1}{2}\leq\frac{\left|x\right|}{L}\leq2n+\frac{3}{2},\end{array}\right.\label{eq:S0e def}\end{align}
where $n=0,1,2,...$ By construction, the function $S_{0}(x;\varepsilon)$ has
the following properties that will be used below: \begin{align}
0\leq S_{0}(x;\varepsilon) & \leq L,\quad\frac{\partial}{\partial x}S_{0}(x;\varepsilon)\leq1;\label{eq:s0prop1}\\
\left|\frac{\partial}{\partial\varepsilon}S_{0}(x;\varepsilon)\right| & \leq\sup_{n,x}\left|\frac{\partial C_{\delta_{n}}(x)}{\partial\varepsilon}\right|<\varepsilon L\leq L.\label{eq:s0prop2}\end{align}
The last property can be derived from Eq.~(\ref{eq:C inequ}). 

The hypersurface $t=S_{0}(\left|\mathbf{x}\right|;\varepsilon)$ is not everywhere
null because of thin spacelike spherical shells in small neighborhoods of $\left|\mathbf{x}\right|=nL$,
$n=0,1,2,$... The 3-volume of these shells can be estimated as\begin{align}
{\cal V}\left[S_{0}(x;\varepsilon)\right] & <\frac{4\pi}{3}\sum_{n=0}^{\infty}\left[\left(nL+\delta_{n}\right)^{3}-\left(nL-\delta_{n}\right)^{3}\right]\nonumber \\
 & <\frac{4\pi}{3}\left(\delta_{0}^{3}+6L^{2}\sum_{n=1}^{\infty}n^{2}\delta_{n}\right)<30\varepsilon L^{3}.\label{eq:bound s0}\end{align}
Therefore the total volume of all the {}``smoothing caps'' can be made negligible
by choosing a sufficiently small $\varepsilon$.

Now we are ready to describe the foliation of the 3+1-dimensional Minkowski
spacetime. The spherically symmetric hypersurface $t=S_{\varepsilon}(\tau,r)$
is constructed by rescaling and shifting the smoothed function $S_{0}(x;\varepsilon)$
defined by Eq.~(\ref{eq:S0e def}). A particular ansatz can be written as\begin{equation}
S_{\varepsilon}\left(\tau,r\right)=\tau+\left(1-h(\tau,r)\right)S_{0}(r;\varepsilon(\tau)),\end{equation}
where $h(\tau,r)$ and $\varepsilon(\tau)$ are suitable functions and $0<h(\tau,r)<1$.
By construction, the surface $S_{\varepsilon}$ is everywhere smooth (infinitely
differentiable) and approaches the {}``skeleton'' surface $S_{0}$ as $h\rightarrow0$,
and we shall set\begin{equation}
h\left(\tau,r\right)=\varepsilon\exp\left(-\frac{r+\tau}{2L}\right),\quad\varepsilon(\tau)=\varepsilon\exp\left(-\frac{\tau}{4L}\right).\label{eq:h eps def}\end{equation}
 It then follows from Eq.~(\ref{eq:s0prop1}) that\begin{align}
0 & \leq S_{\varepsilon}(\tau,r)-\tau\leq(1-h)L<L,\\
\left|\frac{\partial h}{\partial r}\right| & \leq\frac{1}{2L}h,\\
\left|\frac{\partial}{\partial r}S_{\varepsilon}(\tau,r)\right| & \leq1-h+L\left|\frac{\partial h}{\partial r}\right|\leq1-\frac{h}{2}<1.\label{eq:spacelike 1}\end{align}
Therefore the hypersurface $t=S_{\varepsilon}(\tau,r)$ is everywhere spacelike.
The proper 3-volume of this hypersurface is \begin{equation}
{\cal V}\left[S_{\varepsilon}(\tau,r)\right]=4\pi\int_{0}^{\infty}dr\sqrt{^{(3)}g(r)},\end{equation}
where $^{(3)}g$ is the determinant of the induced metric within the surface:\begin{align}
ds^{2} & =\left[1-\left(\frac{\partial S_{\varepsilon}}{\partial r}\right)^{2}\right]dr^{2}+r^{2}d\Omega^{2};\\
^{(3)}g(r) & =\left[1-\left(\frac{\partial S_{\varepsilon}}{\partial r}\right)^{2}\right]r^{4}.\end{align}
We can now divide the hypersurface into the {}``caps,'' i.e.~ranges of $r$
within small neighborhoods of $nL$, $n=0,1,2,...$, and {}``hills'' where
$r$ is outside those neighborhoods. For $r$ within the {}``hills'' we have
$\partial S_{0}/\partial r=\pm1$ and thus the following upper bound holds for
those $r$, \begin{align}
1-\left(\frac{\partial S_{\varepsilon}}{\partial r}\right)^{2}= & \,1-\left(1-h\right)^{2}\left(\frac{\partial S_{0}}{\partial r}\right)^{2}\nonumber \\
 & +2\frac{\partial h}{\partial r}\frac{\partial S_{0}}{\partial r}S_{0}\left(1-h\right)-\left(\frac{\partial h}{\partial r}\right)^{2}S_{0}^{2}\nonumber \\
\leq & \,2h+2\left|\frac{\partial h}{\partial r}\right|S_{0}\leq2h+2\frac{h}{L}L<4h.\label{eq:g ub}\end{align}
So the total volume of the {}``hills'' is less than\begin{equation}
4\pi\int_{0}^{\infty}r^{2}\sqrt{4h(\tau,r)}dr=2048\pi\varepsilon L^{3}\exp\left(-\frac{\tau}{4L}\right).\end{equation}
The total volume of the {}``caps'' is constrained by Eq.~(\ref{eq:bound s0}),
where $\varepsilon$ is now replaced by the function $\varepsilon(\tau)$ defined
by Eq.~(\ref{eq:h eps def}). Therefore an upper bound on the total volume
is found as\begin{align}
{\cal V}\left[S_{\varepsilon}(\tau,r)\right] & <30\varepsilon(\tau)L^{3}+4\pi\int_{0}^{\infty}r^{2}\sqrt{4h(\tau,r)}dr\nonumber \\
 & <6464\varepsilon L^{3}\exp\left(-\frac{\tau}{4L}\right).\label{eq:V bound 0}\end{align}
As we intended, the total 3-volume of the hypersurface $t=S_{\varepsilon}(\tau,r)$
is finite and approaches zero at late times ($\tau\rightarrow\infty$). 

Finally we need to check that the hypersurfaces $t=S_{\varepsilon}(\tau,r)$
are strictly monotonic in $\tau$, i.e.~$\partial S_{\varepsilon}/\partial\tau>0$.
Using Eqs.~(\ref{eq:s0prop2}) and (\ref{eq:h eps def}), we find\begin{align}
\left|\frac{\partial h}{\partial\tau}\right| & \leq\frac{1}{2L}h,\qquad1-\frac{\partial h}{\partial\tau}S_{0}\geq1-\frac{h}{2}>\frac{1}{2};\\
\frac{\partial S_{\varepsilon}}{\partial\tau} & =1-\frac{\partial h}{\partial\tau}S_{0}+\left(1-h\right)\frac{\partial S_{0}(r;\varepsilon)}{\partial\varepsilon}\frac{d\varepsilon(\tau)}{d\tau}\\
 & >\frac{1}{2}-\frac{1}{4L}L>0.\end{align}
Therefore the family of hypersurfaces $S_{\varepsilon}(\tau,r)$ is indeed a
foliation of the entire Minkowski spacetime. This argument completes the construction.

\begin{figure}
\begin{center}\psfrag{L}{$L$}\psfrag{t}{$t$}\psfrag{-L}{$-L$}\psfrag{d}{$l$}\psfrag{-d}{$-l$}

\psfrag{C}{$C(x)$}\psfrag{0}{$0$}\psfrag{x}{$x$}\includegraphics[%
  width=3.2in]{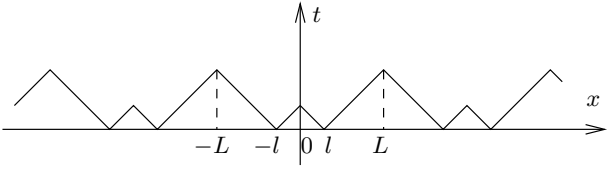}\end{center}

\caption{A typical piecewise-null hypersurface from the foliation in Fig.~\ref{cap:s1}.
\label{cap:SCD}}
\end{figure}

Turning now to the future part of a de Sitter spacetime, say $-H^{-1}<\eta<0$,
we start with the family of piecewise-null hypersurfaces depicted in Fig.~\ref{cap:s1}.
A typical hypersurface having smaller ridges of size $l$ and larger ridges
of size $L$ is shown in Fig.~\ref{cap:SCD}. We denote by $S_{\varepsilon}(x;l,L)$
a smoothed version of this hypersurface constructed using the {}``corner''
functions~(\ref{eq:CDdef}); a cumbersome explicit formula for $S_{\varepsilon}(x;l,L)$
can be written which we omit. One can easily verify the properties\begin{equation}
\left|S_{\varepsilon}(x;l,L)\right|\leq L,\;\left|\frac{\partial S_{\varepsilon}}{\partial x}\right|\leq1,\;\frac{\partial S_{\varepsilon}}{\partial l}\geq0,\;\left|\frac{\partial S_{\varepsilon}}{\partial\varepsilon}\right|\leq L.\label{eq:Sll prop}\end{equation}

The desired family of smooth hypersurfaces $\eta=H_{\varepsilon}(\tau,r)$ is
built, as before, by shifting and rescaling the functions $S_{\varepsilon}(x;l,L)$.
An additional time dependence in the parameters $l$ and $L$ is needed to reproduce
the behavior indicated in Fig.~\ref{cap:s1}. An explicit expression in conformal
coordinates $\left(\eta,r\right)$ can be written as\begin{equation}
\eta=H_{\varepsilon}(\tau,r)=\tau+\left(1-h(\tau,r)\right)S_{\varepsilon(\tau)}(r;l(\tau),L(\tau)),\end{equation}
where we consider the range $-H^{-1}<\tau<0$ (recall that $\tau$ is a conformal
time parameter), the functions $h(\tau,r)$ and $\varepsilon(\tau,r)$ are similar
to those in the Minkowski case, and we define\begin{align}
L(\tau) & =2^{-p(\tau)}L,\quad p(\tau)\equiv-\left\lfloor \log_{2}(-H\tau)\right\rfloor ,\label{eq:Lldef}\\
l(\tau) & =2^{-p(\tau)}L+\frac{1}{2}H\tau L,\label{eq:L2 def}\end{align}
where $\left\lfloor x\right\rfloor $ denotes the integer part of $x$, i.e.~the
algebraically largest integer not exceeding $x$. The definitions~(\ref{eq:Lldef})-(\ref{eq:L2 def})
correspond to Fig.~\ref{cap:s1} and describe a gradual growth of ridges with
size $l(\tau)$. By construction $L(\tau)\leq L\left|H\tau\right|$; more precisely,
during the time range $-H^{-1}2^{-p+1}<\tau\leq-H^{-1}2^{-p}$ we have $L(\tau)=L2^{-p}$
while the ridge size $l(\tau)$ changes from $0$ to $L2^{-p-1}$. At $\tau=-H^{-1}2^{-p}$
the ridge scale $L$ is halved and $l$ is reset to zero. This requires $p\geq0$,
i.e.~$\tau\geq-2H^{-1}$, and the latter inequality holds since we consider
only the range $-H^{-1}\leq\tau<0$. 

The 3-volume of the hypersurface $H_{\varepsilon}(\tau,r)$ is found as\begin{equation}
{\cal V}\left[H_{\varepsilon}(\tau,r)\right]=\frac{4\pi}{H^{3}}\int_{0}^{\infty}r^{2}dr\frac{\sqrt{1-(\partial H_{\varepsilon}/\partial r)^{2}}}{H_{\varepsilon}^{3}}.\end{equation}
 Since the entire hypersurface is bounded in time, namely $H_{\varepsilon}(\tau,r)\leq\frac{1}{2}\tau$,
we may estimate its 3-volume from above using the maximum scale factor $8\left|H\tau\right|^{3}$.
An upper bound on the factor $\sqrt{1-(\partial H_{\varepsilon}/\partial r)^{2}}$
is obtained similarly to Eq.~(\ref{eq:g ub}) since the function $S_{\varepsilon}(r;l,L)$
has properties~(\ref{eq:Sll prop}) similar to those of $S_{0}(x;\varepsilon)$,
so\begin{equation}
1-\left(\frac{\partial H_{\varepsilon}}{\partial r}\right)^{2}<4h(\tau,r).\end{equation}
We now choose $\varepsilon(\tau)$ and $h(\tau,r)$ as before except for the
replacement $\tau\rightarrow-H^{-1}\ln\left(-H\tau\right)$ as appropriate for
the conformal time, \begin{equation}
\varepsilon(\tau)\equiv\varepsilon\exp\left[\frac{\ln\left|H\tau\right|}{4HL}\right],\; h(\tau,r)\equiv\varepsilon\exp\left[\frac{\ln\left|H\tau\right|}{2HL}-\frac{r}{2L}\right].\end{equation}
Then the total volume of the hypersurface is estimated similarly to Eq.~(\ref{eq:V bound 0}),\begin{align}
{\cal V}\left[H_{\varepsilon}\right] & <8\left|H\tau\right|^{-3}\left[30\varepsilon(\tau)L^{3}+4\pi\int_{0}^{\infty}r^{2}\sqrt{4h(r)}dr\right]\nonumber \\
 & <51712\varepsilon L^{3}\left|H\tau\right|^{\frac{1}{4HL}-3}.\end{align}
Choosing the free parameter $L$ (the ridge size) such that $HL<\frac{1}{12}$,
i.e.~having more than $12$ ridges per Hubble length, we find \begin{equation}
\lim_{\tau\rightarrow0}{\cal V}\left[H_{\varepsilon}(\tau,r)\right]=0\end{equation}
as required.

It remains to verify that the hypersurfaces $\eta=H_{\varepsilon}(\tau,r)$
are everywhere spacelike and monotonic in $\tau$. Note that the functions $\varepsilon(\tau)$
and $h(\tau,r)$ satisfy\begin{equation}
\left|\frac{d\varepsilon}{d\tau}\right|=\frac{\varepsilon(\tau)}{4L\left|H\tau\right|}\leq\frac{\varepsilon(\tau)}{4L(\tau)};\;\left|\frac{\partial h}{\partial\tau}\right|=\frac{h(\tau,r)}{2L\left|H\tau\right|}\leq\frac{h}{2L(\tau)}.\end{equation}
The spacelike character is proved by the estimate\begin{equation}
\left|\frac{\partial H_{\varepsilon}}{\partial r}\right|<1\end{equation}
which is derived from Eq.~(\ref{eq:Sll prop}):\begin{align}
\left|\frac{\partial}{\partial r}H_{\varepsilon}(\tau,r)\right| & \leq(1-h)\left|\frac{\partial}{\partial r}S_{\varepsilon}(\tau,r)\right|+\left|\frac{\partial h(\tau,r)}{\partial r}\right|S_{\varepsilon}(\tau,r)\nonumber \\
\leq & 1-h+\frac{h}{2L}L(\tau)\leq1-h+\frac{h}{2}=1-\frac{h}{2}<1.\end{align}
 The monotonicity, $\partial H_{\varepsilon}/\partial\tau>0$, can be shown
using Eq.~(\ref{eq:Sll prop}) and the following inequalities,\begin{align}
\frac{\partial H_{\varepsilon}}{\partial\tau} & =1-\frac{\partial h}{\partial\tau}S_{\varepsilon}+\left(1-h\right)\left[\frac{\partial S_{\varepsilon}}{\partial\varepsilon}\frac{d\varepsilon(\tau)}{d\tau}+\frac{\partial S_{\varepsilon}}{\partial l}\frac{dl(\tau)}{d\tau}\right]\nonumber \\
 & >1-\frac{h}{2L(\tau)}L(\tau)-\frac{\varepsilon}{4L(\tau)}L(\tau)\geq1-\frac{1}{2}-\frac{1}{4}>0.\end{align}
This argument completes the construction of the foliation for the de Sitter
spacetime.

\end{document}